\documentclass[aip,pop,amsmath,amssymb,reprint]{revtex4-1}
\usepackage{graphicx}

\begin{document}

\title[Journal reference: Phys. Plasmas {\bf 21}, 023706 (2014)]{On the metastability of the hexatic phase during the melting of two-dimensional charged particle solids}

\author{Aranka Derzsi}
\affiliation{Institute for Solid State Physics and Optics, Wigner Research Centre for Physics, Hungarian Academy of Sciences, P.O.B. 49, H-1525 Budapest, Hungary}

\author{Anik\'o Zs. Kov\'acs}
\affiliation{Institute for Solid State Physics and Optics, Wigner Research Centre for Physics, Hungarian Academy of Sciences, P.O.B. 49, H-1525 Budapest, Hungary}

\author{Zolt\'an Donk\'o}
\affiliation{Institute for Solid State Physics and Optics, Wigner Research Centre for Physics, Hungarian Academy of Sciences, P.O.B. 49, H-1525 Budapest, Hungary}
\affiliation{Department of Physics, Boston College, Chestnut Hill, MA 02467}

\author{Peter Hartmann}
\affiliation{Institute for Solid State Physics and Optics, Wigner Research Centre for Physics, Hungarian Academy of Sciences, P.O.B. 49, H-1525 Budapest, Hungary}
\affiliation{Department of Physics, Boston College, Chestnut Hill, MA 02467}
\affiliation{Center for Astrophysics, Space Physics and Engineering Research (CASPER), One Bear Place 97310, Baylor University, Waco, TX 76798, USA
}

\date{\today}

\begin{abstract}
For two-dimensional many-particle systems first-order, second-order, single step continuous, as well as two-step continuous (KTHNY-like) melting transitions  have been found in previous studies. Recent computer simulations, using particle numbers in the $\geq 10^5$ range, as well as a few experimental studies, tend to support the two-step scenario, where the solid and liquid phases are separated by a third, so called hexatic phase. We have performed molecular dynamics simulations on Yukawa (Debye-H\"uckel) systems at conditions earlier predicted to belong to the hexatic phase. Our simulation studies on the time needed for the equilibration of the systems conclude that the hexatic phase is metastable and disappears in the limit of long times. We also show that simply increasing the particle number in particle simulations does not necessarily result in more accurate conclusions regarding the existence of the hexatic phase. The increase of the system size has to be accompanied with the increase of the simulation time  to ensure properly thermalized conditions.
\end{abstract}  



\maketitle

\section{Introduction}
The debate about the properties of the melting phase transition of two-dimensional (2D) systems did not lose its intensity over the past several decades. Recent developments in the fabrication of 2D materials \cite{Balleste2011} simultaneously seek for, and may provide clarification of the details of the transition. A milestone, and still the most widely accepted theory available, is the Kosterlitz-Thouless-Halperin-Nelson-Young (KTHNY) picture \cite{Halperin1978,*Halperin_err}. In the underlying physical process two separate, continuous transitions can be distinguished, as the solid transforms into a liquid in quasi-equilibrium steps by slow heating. During the first stage the translational (positional) order vanishes, while in the second stage the orientational order decays. All this is mediated by the unbinding of (i) dislocation pairs into individual dislocations, and (ii) dislocations into point defects \cite{Strandburg1988}. The strength of this theory consists in its compatibility with the Mermin-Wagner theorem that forbids the existence of exact long range positional order in 2D for a wide range of pair potentials, at finite temperatures \cite{Mermin1968}. The most criticized weakness of it, however, is that it assumes a dilute, unstructured distribution of the lattice defects, which is in contradiction with observations, where the alignment and accumulation of dislocations into small angle domain walls was found \cite{Groma2007}.

Since the birth of the KTHNY theory, the examination of its validity for systems with different pair interactions has been in focus. Investigations started with hard-sphere (disk), Lennard-Jones, and Coulomb systems. More recently, systems characterized by dipole-dipole and Debye-H\"uckel (screened Coulomb or Yukawa) inter-particle interactions became important due to the significant advances achieved in the field of colloid suspensions \cite{Ebert2009} and dusty plasmas \cite{Morfill2009}.

To illustrate the incongruity of both experimental and simulation results that had accumulated over the last three decades on investigations of classical single-layer (2D) many-body systems, we list a few examples:

\begin{itemize}
\item First order phase transition to exist was reported for Lennard-Jones systems \cite{Farid1981,Kleinert1983,Kleinert1989} and hard-disk systems \cite{Ryzhov1995,Strand1992}, for the phase-field-crystal (PFC) model \cite{Grana2010,Grana2011}, as well as for Coulomb and dipole systems \cite{Kali1981} in simulations, and in experiments with halomethanes and haloethanes physisorbed on exfoliated graphite \cite{Knorr1998}, as well as in experiments on a quasi-two-dimensional suspension of uncharged silica spheres \cite{Karnch2000}.
\item Second order (or single step continuous) transition was found in dusty plasma experiments \cite{Melzer1996,Takahashi1999,Goree2001,Sheridan2008,Nosenko2009}, for a hard-disk system  \cite{Fernandez1997}, electro-hydrodynamicly excited colloidal suspensions \cite{Dutcher13}, as well as for Coulomb \cite{Kleinert1989} and Yukawa \cite{Harti2007} systems.
\item KTHNY-like transition was reported in a dusty plasma experiment \cite{LinI2001} and related numerical simulations \cite{Vaulina13}, for the harmonic lattice model \cite{Kleinert2006}, in experiments and simulations of colloidal suspensions \cite{Kusner1995,Zahn1999,Zahn2000,Qi2006,Keim2007,Peng2010,Bordin2010,SvenHex13}, for Lennard-Jones \cite{Shiba2009,Gribova11}, Yukawa \cite{Sheridan2009,Qi2010}, hard disk \cite{Binder2002,Mak2006,Bernard2011,Engel13}, dipole-dipole \cite{Lin2006,Schock13}, Gaussian-core \cite{Santi11}, and electron systems \cite{Ryzhov1995,Muto1999,He2003}, as well as for a system with $r^{-12}$ repulsive pair potential \cite{Bagchi1996}, weakly softened core \cite{Santi12}, and  for vortices in a W-based superconducting thin film \cite{Nature2009}.
\end{itemize}

The effect of the range of the potential on two-dimensional melting was studied in \cite{Lee2008} for a wide range of Morse potentials. It has been shown, that extended-ranged interatomic potentials are important for the formation of a ``stable'' hexatic phase. Similar conclusion was drawn in \cite{Pomirchi2001} for modified hard-disk potentials. The effect of the dimensionality (deviation from the mathematically perfect 2D plane) on the hexatic phase was discussed for Lennard-Jones systems in \cite{Gribova11}. It was found, that an intermediate hexatic phase could only be observed in a monolayer of particles confined such that the fluctuations in the positions perpendicular to the particle layer was less than 0.15 particle diameters.

The timeline of the results listed above shows a general trend: in earlier studies, first or second order phase transitions were identified in particle simulations, but subsequently, as the computational power increased with time, since approximately the year of 2000, particle based numerical studies became in favor of the KTHNY theory. A possible resolution of the ongoing debate is given in \cite{Chen1995}, where extensive Monte Carlo simulations of 2D Lennard-Jones systems have revealed the metastable nature of the hexatic phase. This seems to support PFC simulations \cite{Grana2010,Grana2011} operating on the diffusive time-scale (averaging out single particle oscillations), which is significantly longer, than what Monte Carlo (MC) or Molecular Dynamics (MD) methods can cover.

In this paper we will show that the observation of the hexatic phase is strongly linked with the thermodynamic equilibration of the systems. The necessary equilibration time, in turn, strongly depends on the measured quantity of interest. Local, or single particle properties can equilibrate very rapidly, while long-range, or collective relaxations usually take significantly longer. We find, consequently, that monitoring the velocity distribution function alone to verify the equilibration of the system is insufficient. The idea, that numerical simulations may have related equilibration issues (called as kinetic bottlenecks) was raised already in 1993 in \cite{Naidoo93}.

\section{Molecular dynamics simulations}

We have performed extensive microcanonical MD simulations \cite{MD} in the close vicinity of the expected solid-liquid phase transition temperature, $T_m$, for repulsive screened Coulomb (also called Yukawa or Debye-H\"uckel) pair-potential with the potential energy in form of 
\begin{equation}
\Phi(r) = \frac{q^2}{4\pi\varepsilon_0} \frac{\exp(-r/\lambda_D)}{r},
\end{equation}
where $\lambda_D$ is the Debye screening length, $q$ is the electric charge of the particles, and $\varepsilon_0$ is the vacuum permittivity. To characterize the screening we use the dimensionless screening parameter $\kappa=a/\lambda_D$, where $a=1/\sqrt{\pi n}$ is the Wigner-Seitz radius, and $n$ is the particle density. This model potential was chosen because of its relevance to several experimental systems consisting of electrically charged particles, like dusty plasmas, charged colloidal suspensions, and electrolytes. Here we show results obtained for $\kappa=2$. Our earlier studies \cite{Harti2007,Torben11} identified the melting transition (without clarifying its nature) to take place around the Coulomb coupling parameter 
\begin{equation}
\Gamma_m = \frac{q^2}{4\pi\varepsilon_0} \frac{1}{a k_{\rm B} T_m} = 414 \pm 4
\end{equation}
for this strength of screening.

Time is measured in units of the nominal 2D plasma oscillation period with
\begin{equation}
\omega^2_0=\frac{n q^2}{2\varepsilon_0 m a},
\end{equation}
where $m$ is the mass of a particle. Our simulations are initialized by placing $N$ particles (in the range of 1,920 to 740,000) into a rectangular simulation cell that has periodic boundary conditions. The particles are released from hexagonal lattice positions, with initial velocities randomly sampled from a predefined distribution. At the initial stage, which has a duration $t_0$ (thermalization time), the system is thermostated by applying the velocity back-scaling method (to follow the usual approach used in many previous studies) to reach near-equilibrium state at the desired (kinetic) temperature. Data collection starts only after this initial stage and runs for a time period $t_m$ (measurement time) without any additional thermostation. 

To characterize the level of equilibration we study the time and system size dependence of the following quantities: 
\begin{itemize}
\item momenta of the velocity distribution function, $f(v)$, 
\item the configurational temperature, $T_{\rm conf}$ \cite{Rugh97}, and 
\item the long-range decay of the $g(r)$ pair-correlation and $g_6(r)$ bond-angle correlation functions \cite{Halperin1978,Halperin_err,Qi2006}.
\end{itemize}
While in the case of the first two quantities $t_0=0$, in the simulations targeting the correlation functions, $t_0$ is varied over a wide range and the measurement time is chosen to be $t_m \ll t_0$ to avoid significant changes (due to ongoing equilibration) during the measurement.

\subsection{Velocity momenta}

Using the Maxwell-Boltzmann assumption for the velocity distribution in thermal equilibrium in the form
\begin{equation}
f(v)=\frac2\tau v\exp(-v^2/\tau),
\end{equation}
where $\tau=2kT/m$, in two-dimensions the first four velocity moments are: 
\begin{eqnarray}
\langle v\rangle &=&\frac12\sqrt{\pi \tau} \\
\langle v^2\rangle &=& \tau \nonumber \\
\langle v^3 \rangle &=& \frac{3\sqrt{\pi}}{4}\tau^{3/2} \nonumber \\
\langle v^4\rangle &=& 2\tau^2 \nonumber
\end{eqnarray}

\begin{figure}[htb]
\centering
\includegraphics[width=8cm]{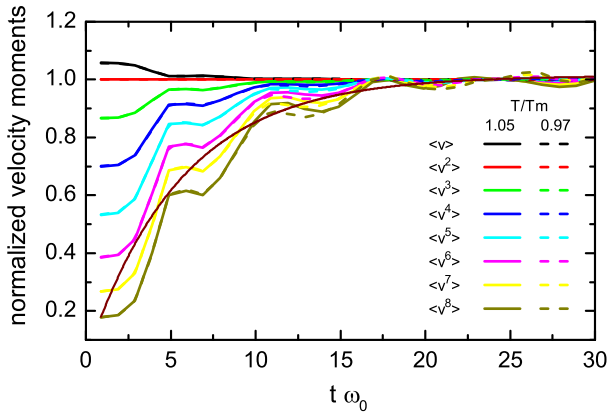}
\caption{\label{fig:vmom} 
(color online) Moments of the computed velocity distribution functions relative to the theoretical equilibrium values vs. simulation time at temperatures slightly above (full  lines) and below (dashed lines) the melting point, $T_m$. The dashed lines are mostly hidden behind the full lines, indicating a low sensitivity on the temperature. The dark red curve shows functional fit in the form $y=y_0+Ae^{-t/t_r}$ to $\langle v^7\rangle$. $N=184,400$.}
\end{figure}

To measure the relaxation time of the velocity distribution function we have performed MD simulations with particle numbers $N=184,400$ and $N=7520$, with initial velocity components ($x$ and $y$) sampled from a uniform distribution between $-\sqrt{2k_{\rm B}T/m}$ and $\sqrt{2k_{\rm B}T/m}$, in order to start with the desired average kinetic energy, but being far from equilibrium. Figure \ref{fig:vmom} shows the time evolution of the first eight velocity moments normalized with their theoretical equilibrium values. As already mentioned, the initial conditions are far from the equilibrium configuration (perfect lattice position and non-thermal velocity distribution).

We can observe, that the velocity momenta have initial values very different from the expected Maxwell-Boltzmann equilibrium distribution. The values approach the equilibrium value asymptotically with regular oscillations. These oscillations (or fluctuations) are typical for microcanonical MD simulations, where the total energy of the system is constant, while there is a permanent exchange of potential and kinetic energies. The relaxation time can be found by fitting the curves with an exponential asymptotic formula in the form $y=y_0+Ae^{-t/t_r}$. We find, that the relaxation of the velocity distribution can be characterized by a short relaxation time of $t_r\approx 5.5/\omega_0$, and this is independent of system size and temperature in the vicinity of the melting point. 

\subsection{Configurational temperature}

\begin{figure}[htb]
\centering
\includegraphics[width=8cm]{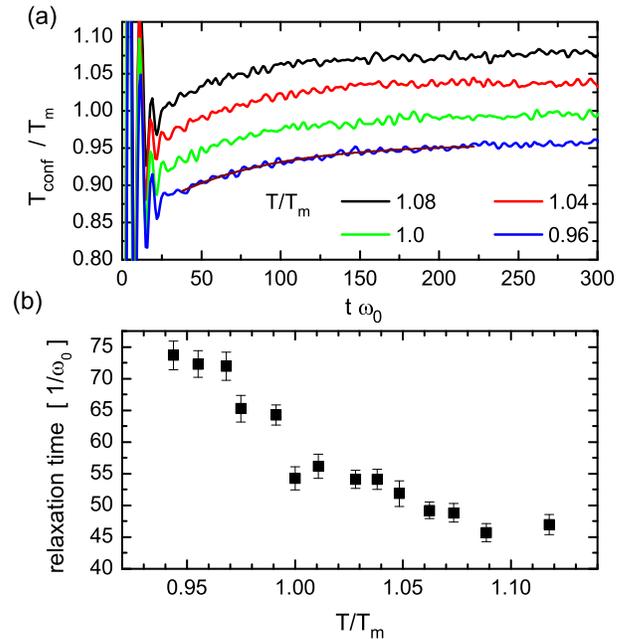}
\caption{\label{fig:tconf} 
(color online) (a) Time evolution of the configurational temperature $T_{\rm conf}$ for different kinetic temperatures $T$. (b) Relaxation time vs. kinetic temperature. ($N=184,400$).}
\end{figure}

In 1997, Rugh \cite{Rugh97} pointed out that the temperature can also be expressed as ensemble average over geometrical and dynamical quantities and derived the formula for the configurational temperature: 
\begin{equation}
kT_{\rm conf}=-\langle\sum_{i=1}^{N} F_i^2\rangle / \langle\sum_{i=1}^{N} \nabla {\bf F}_i\rangle,
\end{equation}
where ${\bf F}_i=-\sum_{j\ne i}^{N}\nabla \Phi(r_{ij})$. As the central quantity in this expression is the inter-particle force acting on each particle, in case of finite range interactions (like the Yukawa potential), the configurational temperature is sensitive on the local environment within this range. Simulations were performed for a series of particle numbers between $N=1920$ and $N=740,000$ with initial velocities sampled from Maxwellian distribution. Figure~\ref{fig:tconf}(a) shows examples from runs with $N=184,400$ for the time evolutions, while fig.~\ref{fig:tconf}(b) presents relaxation time data computed (similarly as above) for different kinetic temperatures.

We observe relaxation times about an order of magnitude longer ($t_r \approx 55/ \omega_0$) compared to the velocity distribution, and a strong temperature dependence in the vicinity of the melting point. No significant system size dependence was found.

\subsection{Correlation functions}

\begin{figure}[htb]
\centering
\includegraphics[width=8cm]{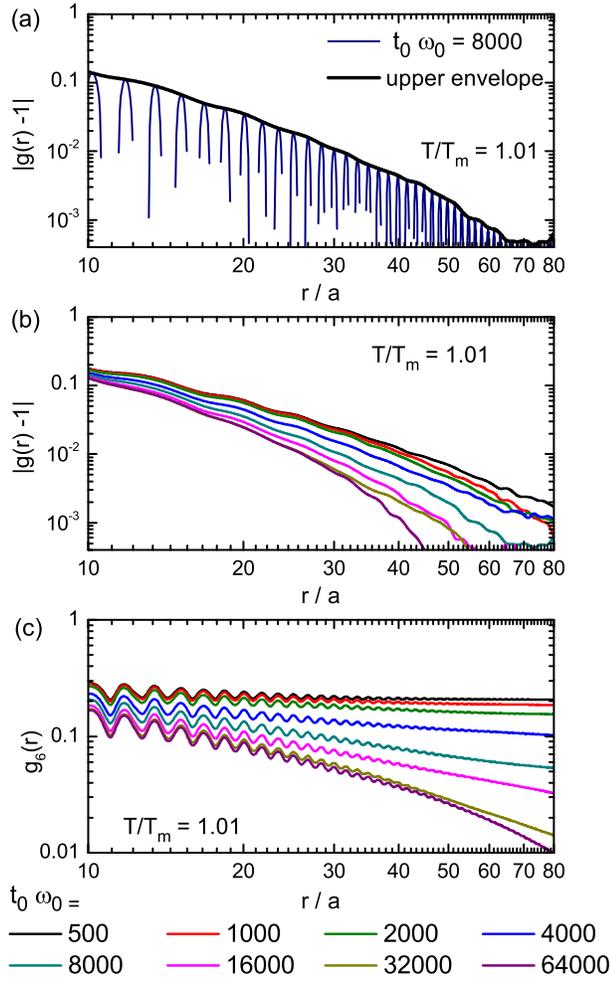}
\caption{\label{fig:corr} 
(color online) Log--log plots of (a) an example of $g(r)-1$ pair correlation function with its upper envelope, (b) a series of envelope curves of pair correlation functions, (c) $g_6(r)$ bond-order correlation functions measured after letting the systems equilibrate for various times indicated, $t_0$, at a temperature 1 percent above the melting point. The systems consisted of  $N=104,400$ particles, the data acquisition took $t_m = 500/\omega_0$ and started after $t_0$ has elapsed.} 
\end{figure}

\begin{figure}[htb]
\centering
\includegraphics[width=8cm]{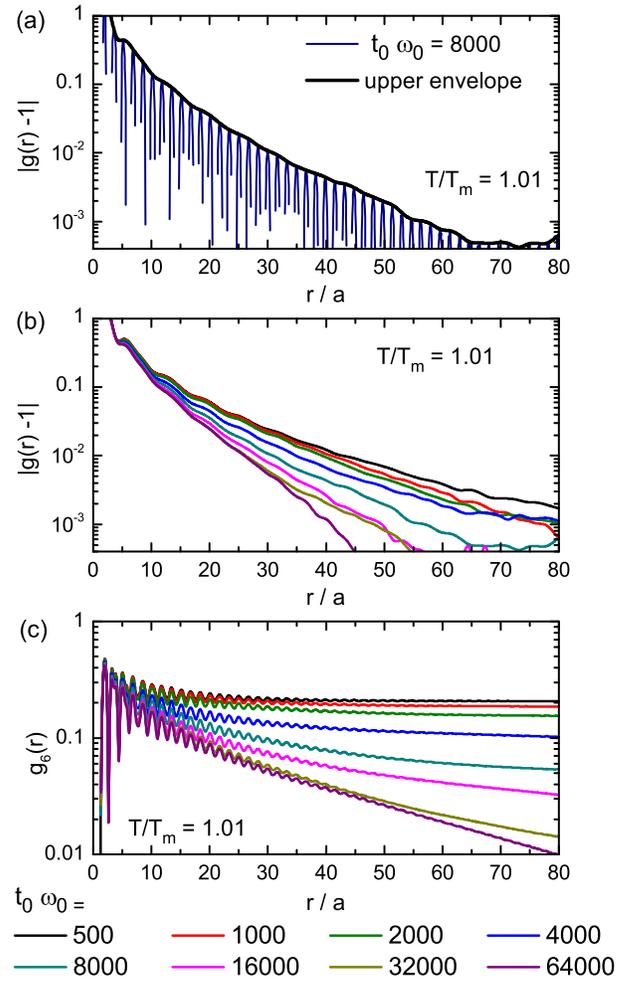}
\caption{\label{fig:corr2} 
(color online) Same as fig~\ref{fig:corr} with semi-logarithmic scales.} 
\end{figure}

The central property used to identify the hexatic phase is traditionally the long-range behavior of the pair-, and bond-order correlation functions, $g(r)$ and $g_6(r)$, respectively \cite{Halperin1978,Halperin_err,Qi2006}. To be able to compute correlations at large distances, one naturally has to use large particle numbers, otherwise the periodic boundary conditions introduce artificial correlation peaks. This trivial constraint led to investigations of larger and larger systems by different groups. Figures~\ref{fig:corr} and \ref{fig:corr2} show correlation functions for systems consisting of $N=104,400$ particles, for a set of increasing equilibration times provided to the systems before performing the data collection. 

We can observe a clear long-time evolution of the correlation functions. On the double-logarithmic plot the $g(r)$ pair-correlation functions show already at early times a long-range decay, which is faster than power-law [fig.~\ref{fig:corr}(a,b)], while the $g_6(r)$ orientational correlations smooth out to near perfect straight lines [fig.~\ref{fig:corr}(c)], representing power-law type decay for relatively long times. On the semi-logarithmic graphs all the $g(r)$ functions have almost straight upper envelopes [fig.~\ref{fig:corr2}(a,b)] in the intermediate distance range $10 < r/a < 70$, where the statistical noise is still negligible. This indicates almost pure exponential decay, although the characteristic decay distance does decrease with increasing simulation time. On the other hand, it is only the last $g_6(r)$ orientational correlation function, belonging to the longest simulation, that shows linear apparent asymptote on the semi-logarithmic scale [fig.~\ref{fig:corr2}(c)], representing a clear exponential decay, meaning the lack of long range order. To conclude these observations: in short simulations we observe short-range positional and quasi-long-range orientational order, signatures of the hexatic phase, which, however vanish if we provide the system longer time for equilibration. As a consequence, in the case we would stop the simulation at, e.g., $t_0=8000/\omega_0$ (which already means simulation time-steps in the order of $10^5$, as plasma oscillations have to be resolved smoothly) we may identify the system to be in the hexatic phase, exactly as shown in \cite{Qi10}, which, however is not the true equilibrium configuration.

In addition, as the accessible length scale strongly depends on the system size (typically less than 1/3 of the side length of the simulation box), smaller systems apparently equilibrate faster. We have found $t_0 \approx 4000/\omega_0$ to be sufficient to reach equilibrium for a system of $N=1920$ particles, while $t_0 \approx 64,000/\omega_0$ was needed for $N=104,400$. 
 
To verify, that the observed slowdown of relaxation is not an artifact of the applied microcanonical (constant $NVE$) simulation, we have implemented the computationally much more demanding, but in principle for phase transition studies better suited isothermal-isobaric (constant $NPT$) molecular dynamics scheme \cite{NPT}. Although the $NPT$ simulations were performed for much smaller systems ($N=1020$), limiting the calculation of the correlation functions to a shorter range and resulting in higher noise levels, the same long-time tendency of decaying long-range correlations could be identified as already shown with the computationally much more efficient microcanonical simulations.

\section{Conclusions}

During the equilibration of an interacting charged many-particle system we have identified three different stages of relaxation:
\begin{itemize}
\item The velocity distribution does approach the Maxwellian distribution within a few plasma oscillation cycles. In the close vicinity of the melting transition the speed of this process is found to be independent of temperature and system size.
\item Compared to the velocity distribution function, the configurational temperature (determined by the local neighborhood within the range of the inter-particle interaction potential) relaxes at time scales about an order of magnitude longer for our systems. The relaxation time scale is not sensitive to the system size, but has a strong dependence on the temperature.
\item The equilibration of the long range correlations is significantly slower compared to the above quantities, and depends strongly on the systems size (larger systems need longer time to equilibrate). 
\end{itemize}

From this study we can conclude, that increasing the system size in particle simulations alone can be insufficient and can result in misleading conclusions, as the length of the equilibration period also plays a crucial role in building up or destroying correlations.

In the vast majority of the earlier numerical studies on charged particle ensembles (as listed in the Introduction) no simulation time is specified, given to the system to equilibrate before the actual measurement were performed, neither is the method of characterizing the quality of the equilibrium described. Based on these results, we suspect, that the rapidly increasing computational resources in the first decade of the 21st century beguiled increasing the system sizes in particle simulations without increasing the length of the simulated time intervals. In the majority of these studies the systems may got stuck in the metastable hexatic phase, instead of settling in the true equilibrium configuration. 

\acknowledgements
We appreciate useful discussions with Profs. Gabor J. Kalman, L\'aszl\'o Gr\'an\'asy, and Andr\'as S\"ut\H{o}. This research has been supported by the OTKA Grants NN-103150 and K-105476.

\nocite{*}
%

\end{document}